\documentclass[runningheads]{llncs}
\usepackage[T1]{fontenc}
\usepackage{graphicx}
\usepackage{todonotes}
\presetkeys%
    {todonotes}%
    {inline,backgroundcolor=yellow}{}
\usepackage{blindtext}

\usepackage{stfloats}
\usepackage{enumitem}
\setitemize{parsep=4pt}

\PassOptionsToPackage{hyphens}{url}
\usepackage[hidelinks]{hyperref}

\colorlet{punct}{red!60!black}
\definecolor{background}{HTML}{EEEEEE}
\definecolor{delim}{RGB}{20,105,176}
\colorlet{numb}{magenta!60!black}
\usepackage{listings}
\lstdefinelanguage{json}{
    basicstyle=\footnotesize\ttfamily,
    numbers=left,
    numberstyle=\scriptsize,
    stepnumber=1,
    numbersep=8pt,
    showstringspaces=false,
    breaklines=true,
    frame=lines,
    backgroundcolor=\color{background},
    literate=
     *{0}{{{\color{numb}0}}}{1}
      {1}{{{\color{numb}1}}}{1}
      {2}{{{\color{numb}2}}}{1}
      {3}{{{\color{numb}3}}}{1}
      {4}{{{\color{numb}4}}}{1}
      {5}{{{\color{numb}5}}}{1}
      {6}{{{\color{numb}6}}}{1}
      {7}{{{\color{numb}7}}}{1}
      {8}{{{\color{numb}8}}}{1}
      {9}{{{\color{numb}9}}}{1}
      {:}{{{\color{punct}{:}}}}{1}
      {,}{{{\color{punct}{,}}}}{1}
      {\{}{{{\color{delim}{\{}}}}{1}
      {\}}{{{\color{delim}{\}}}}}{1}
      {[}{{{\color{delim}{[}}}}{1}
      {]}{{{\color{delim}{]}}}}{1},
}

\usepackage{color}

\begin{document}
\title{Container Orchestration Patterns\\ for Optimizing Resource Use}
\author{Diogo Maia\inst{1} \and
Filipe F. Correia\inst{1} \and
André Restivo\inst{2} \and
Paulo G. G. Queiroz \inst{1}\inst{3}}
\authorrunning{Maia et al.}
\institute{INESC TEC, Faculty of Engineering, University of Porto, Portugal\\
\email{\{up201904974,filipe.correia\}@fe.up.pt} \and
LIACC, Faculty of Engineering, University of Porto, Portugal\\
\email{arestivo@fe.up.pt} \and
Federal University of the Semi-Arid Region, Brazil\\
\email{pgabriel@ufersa.edu.br}}

\maketitle              % typeset the header of the contribution
\begin{abstract}
Service-based architectures provide substantial benefits, yet service orchestration remains a challenge, particularly for newcomers. While various resources on orchestration techniques exist, they often lack clarity and standardization, making best practices difficult to implement and limiting their adoption within the software industry. 

To address this gap, we analyzed existing literature and tools to identify common orchestration practices. Based on our findings, we define three key orchestration resource optimization patterns: {\sc Preemptive Scheduling}, {\sc Service Balancing}, and {\sc Garbage Collection}. {\sc Preemptive Scheduling} allows the allocation of sufficient resources for services of higher priority in stressful situations, while {\sc Service Balancing} enables a restructuring of the nodes to allow better resource usage. To end, {\sc Garbage Collection} creates cleanup mechanisms to better understand the system's resource usage and optimize it. These patterns serve as foundational elements for improving orchestration practices and fostering broader adoption in service-based architectures.

\end{abstract}

\section{Introduction}

The rapid advancement in microservices-based applications~\cite{jetbrains_survey,de2023charm}, along with the extensive adoption of container technologies for software packaging, has led to a significant increase in the use of orchestration tools~\cite{new_relic}. The term "orchestration" can have different meanings depending on the context. To avoid confusion and ensure alignment throughout this paper, we include specific descriptors to clearly define our focus. Here, we focus on orchestration in the context of managing service-based applications, which closely relates to container orchestration.
Today, containers are one of the most common methods for packaging and hosting services. Therefore, in this study, orchestration is defined as:

\begin{quote}
    \textit{Container orchestration allows cloud and application providers to define how to select, to deploy, to monitor, and to dynamically control the configuration of multi-container packaged applications in the cloud}~\cite{casalicchio2019container}
\end{quote}

This paper focuses on orchestration tasks during deployment and run-time, including scheduling, scaling, monitoring, and fault tolerance. As service-based containerized systems continue to proliferate, the corresponding need for orchestration has prompted the emergence and broad adoption of several best practices. These practices have shaped the evolution of orchestration tools as developers implement new features to address emerging challenges. However, these practices are not always well-documented or widely understood, leaving practitioners without clear guidance on their application, timing, or intended outcomes.

To bridge this gap, these practices can be documented as design patterns. Although many patterns have been specified for service-based and cloud-native systems~\cite{cloudpatterns2021brown,sousa_engineering_2017,sousa2015patterns,Sousa2022ASO,service_mesh,sousa2018engineering,albuquerque2022proactive,albuquerque2023deployment,albuquerque2024logging}, only a limited number address orchestration. This paper continues our work~\cite{configpatterns} on defining additional orchestration patterns, contributing three patterns related to previously defined patterns, which aid in optimizing system resources.

This article begins by detailing the research methodology used to uncover the patterns and practices discussed. Subsequently, the paper introduces a pattern map derived from the findings. Following this, the three newly identified patterns are thoroughly explained. The paper concludes with a discussion of the contributions made and potential directions for future research.

\section{Research Process}

These design patterns are derived from two primary sources: existing literature and orchestration tools, including orchestrators, add-ons, and libraries. These sources were critical to gaining an in-depth understanding of service orchestration, identifying research gaps, and formalizing best practices into patterns.

The literature review focused on exploring existing research on orchestration and identifying established patterns. In addition, gray literature, such as blogs and forums, was examined to uncover industry practices and address gaps in formal research.

A feature analysis of orchestration tools was conducted to complement the literature findings. This analysis covered Kubernetes~\cite{kubernetes}, Docker Swarm~\cite{docker-swarm}, and Mesos~\cite{mesos}, chosen due to their prominence in the literature and their foundational design (unlike tools built on top of other orchestrators, such as CloudFoundry~\cite{cloudfoundry}). These tools' documentation, repositories, and pull requests were analyzed to extract further insights.

The research identified 29 patterns related to container orchestration and 8 promising practices that have not yet been formalized. These findings were organized into a pattern map that highlights connections between patterns and how some address the consequences of others. The full process and results are detailed in Maia's master's thesis~\cite{maia2024thesis}. This article focuses on 3 of the 8 promising practices that were developed into patterns: {\sc Preemptive Scheduling}, {\sc Service Balancing}, and {\sc Garbage Collection}.

\section{About the Patterns}\label{sec:patterns}

\begin{figure}[]
    \centering
    \includegraphics[scale=0.8]{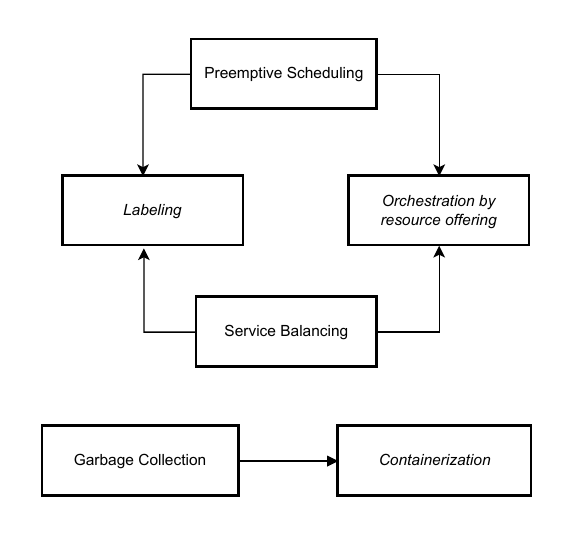}
    \caption{Pattern Map of the Defined Patterns and Direct Connections}
    \label{fig:pat_map1}
\end{figure}

The pattern map shown in Figure \ref{fig:pat_map1} is a visual representation of the 3 patterns we defined from the research process explained previously and their direct connections, as in patterns that can be directly used in the implementation or complement each other. These connections are represented by a directed line from one pattern to another.

The patterns shown in italics are those already established in the literature. They are:
\begin{itemize}
    \item {\sc Labeling} refers to the use of metadata tags (labels) to manage, group, and select containers or other resources in a flexible and scalable way ~\cite{configpatterns}
    \item {\sc Orchestration by resource offering} allows every container/job to have the required resources to function properly, ensuring the stable run of the services~\cite{sousa2015patterns}.
    \item {\sc Containerization} encapsulates a service's files and dependencies, enabling the deployment of the service within a self-contained and isolated environment~\cite{sousa2015patterns}. 
\end{itemize}

\section{Preemptive Scheduling}

\subsection{Context}
A distributed system is comprised of a cluster of multiple nodes. Some nodes run services while others are empty nodes, awaiting some services to be scheduled for them. These nodes are comprised of a group of containers, which will host the services on them. Several services are to run in nodes with various requirements and specifications. These requirements and specifications are directly connected to the resources or properties of the node (all known by the developers), as the match of both will allow the service to run as intended by the developers. An example of a service requirement is the geographical location where the containers in a node are running. The developer wants a certain service to be hosted in containers of certain areas, such as Europe, so the system will try to schedule the service to a node where the containers are running in each zone.

All the system's services are not of the same importance. Some services are essential for the system's well-being, and without them, the system collapses. Other services are still necessary but not crucial for the service's runtime, making some system functionalities not work, but allowing the system as a whole to function in a minimum viable way. Some other services aren't as essential as others, such as being able to be shut down and the system working as intended for a user. 

There are scenarios where crucial services need some node resources that can or may not already be assigned to some other service, or less important services try to take resources from the more critical services. Not managing situations like this can lead to essential services being starved for resources or a lack of optimization on resource usage and costs.
 
\subsection{Example}

\begin{figure*}[]
    \centering
    \includegraphics[width=\textwidth]{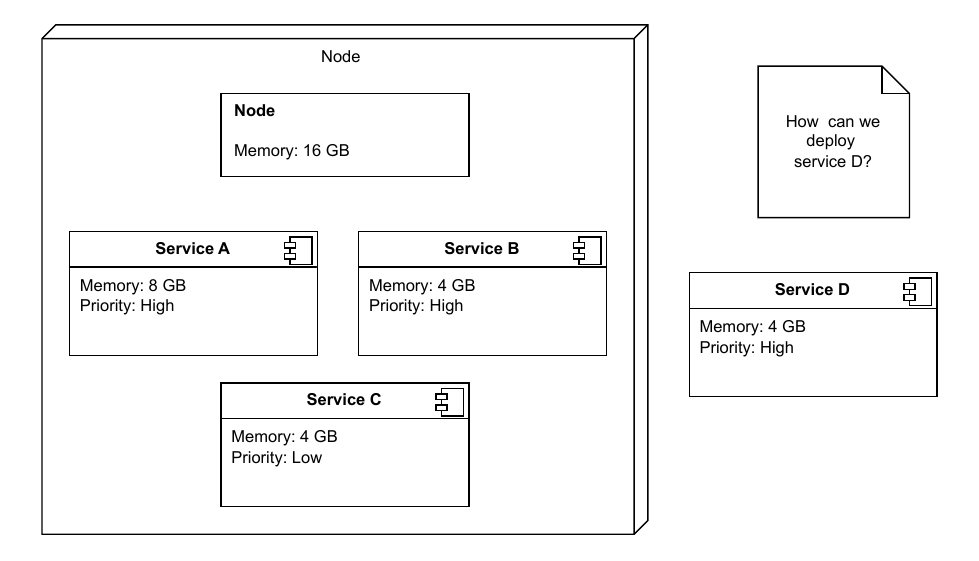}
    \caption{Preemptive Scheduling - Visual Representation of the Example}
    \label{fig:ps_1}
\end{figure*}

As we can see in Figure \ref{fig:ps_1}, a system built using microservices architecture comprises 3 different services: a frontend web server (A), an API server (B), and a data processing server (C). In terms of importance for the system's developers, the first two services are of higher importance since they are directly connected to the website's well-being, and the website's availability is one of the developers' biggest concerns, since they are concerned about costs. The last service is essential to the system as a whole, but it is a background task that, in case of its shutdown, does not directly affect the website's well-being.

This system is hosted on one node. Typically, the three services are deployed and scheduled at the same node, leaving no free node resources. At specific points, the developers decide to deploy a new temporary service in charge of a critical task (Service D), more important than the data processing server. However, there are no resources for it in the node.
 
\subsection{Problem}

How can we ensure that crucial services are getting priority when it comes to their requirements compared to other less important services?
 
\subsubsection{Forces}

\begin{itemize}
    \item \textbf{Availability/ Resilience} - Ensuring crucial services have the resources they need during runtime allows the system to be available at all intended times. By not doing it, the system can go down, which can affect the developer's productivity and users' experience of the system;
    \item \textbf{Consistency} - By guaranteeing crucial service requirements, the developers allow the service to run as efficiently as possible. This allows the crucial part system to behave as intended, even in spikes of demand or costly operations; 
    \item \textbf{Resource/Cost Management} - There is a cost associated with the resources that the developers can own and use. The lower the costs, the fewer resources the system will have to run with. In case the resources are low, there may be a need to take resources from one service to give to another. If the resources are high, the developers can try to guarantee resources for each service;
    \item \textbf{Hierarchy} - The system can have all services with the same or very similar levels of importance;
    \item \textbf{Adaptability} - The priority of a service can change according to the business needs and system state;
    \item \textbf{Customization} - Not all services, if they have priority, have the same degree of priority. A service can have priority over all other services, but a service may also be more important than one service while being less important than another.
    
\end{itemize}

\subsection{Solution}
 
\textit{Define priority levels for each service, and when a service with a higher priority needs the resources of a lower priority service, evict it and try to reschedule it.}

\vspace{0.1cm}

To effectively manage resources for high-priority services, the system can evict lower-priority services from a node, reallocating resources to crucial services. Once evicted, the lower-priority service is sent back to the scheduler for rescheduling when resources become available.

The first step in implementing this pattern is to define service priorities. Not all services should be treated as critical. Priority should be reserved for services essential to business operations, such as user-facing applications or key API servers that require high availability. Over-assigning priority to many services can lead to excessive preemption, disrupting less critical but necessary operations. The priority of each service must be clearly defined in its configuration and should be adjustable based on business needs or other factors. These adjustments can be made manually or via an API.

To facilitate effective scheduling, it's important to establish precise priority levels. This can be as simple as binary levels—priority or non-priority—but it may also involve more granular distinctions, such as numerical levels from 1 to 10. In this system, a priority level of 1 would represent the highest priority, ensuring the service receives all necessary resources. In contrast, a level 10 would indicate a low-priority service that can be delayed. Alternatively, priority levels could be represented by letters or codenames. Regardless of how the levels are defined, they must be standardized across all services to ensure consistent comparisons during scheduling.

During the scheduling process, tasks should be organized into a priority queue, with higher-priority tasks placed at the front. This approach ensures that crucial services are started first and minimizes the need to evict lower-priority services, which can introduce time-related overhead.

When scheduling, the system should first attempt to find a node that meets the service's resource requirements without considering priority levels. If no suitable node is available, the scheduler compares the priority levels of services running on potential nodes. If a lower-priority service is identified, it is evicted, and the high-priority service is deployed. The evicted service is then re-added to the scheduler's priority queue. If no suitable node is found because all nodes run services of equal or higher priority, the high-priority service will be scheduled later.

In the last paragraph, we discuss a default way to deal with low-priority services after eviction. However, this can be customized by the developers as they believe it should better serve the service architecture. For example, suppose the service evicted runs in a certain time slot. In that case, the developer can decide not to try to reschedule the service after it is evicted, knowing it will be deployed and scheduled at a later time. Another example of dealing with the evicted service can be delegating its canceled work to another instance of the service currently running, improving the system's fault tolerance and guaranteeing the work is done. This adds a layer of customization for developers, leading to a better system as a whole.

\subsection{Implementation}
Kubernetes has an innate preemption when scheduling pods to nodes. According to their documentation\cite{kubernetes_preemption}, this is achieved by defining one or more \textit{PriorityClass} objects, which assign a specific priority level to pods within the system.
 
\begin{lstlisting}[language=json,caption= Example of PriorityClass object creation in Kubernetes]
    apiVersion: scheduling.k8s.io/v1
    kind: PriorityClass
    metadata:
      name: high-priority
    value: 1000000
    globalDefault: false
    description: "this priority class should be used for XYZ service pods only."
\end{lstlisting}

Kubernetes allows for creating priority objects with a high level of customization. The following are the main attributes that define this object:
    \begin{itemize}
        \item \textbf{name}: the name the developer wants to give to the priority level. It can either be a way to describe just the level by itself, like in this case, or add some usage context;
        \item \textbf{value}: In Kubernetes, value can be any 32-bit integer value smaller than or equal to 1 billion. The higher the level, the higher the priority of the pod;
        \item \textbf{globalDefault}: Only one Kubernetes \textit{Priority Class} can be defined as the global default (by having this property defined as "true". When true, all the other pods in the 
        system without any \textit{Priority Class} attached will have this priority level; 
        \item \textbf{description}: The developers can, in this property, define information regarding the usage of this object.
    \end{itemize}

By defining the priority levels as objects, Kubernetes incites developers to assign certain \textit{Priority Classes} to groups of Pods with the same priority and context. For example, creating a \textit{PriorityClass} for data processing services allows them to not only assign the same priority to these services but also, in a moment where there is a business need to change them, instead of changing one by one, developers only need to change the \textit{PriorityClass} object. 

In the Pod configurations, the developers can define the priority level by specifying the property \textit{priorityClassName} as the name property of the intended \textit{PriorityClass} object.

\begin{lstlisting}[language=json,caption= Example of defining the PriorityClass of a Pod in Kubernetes]
    apiVersion: v1
kind: Pod
metadata:
  name: nginx
  labels:
    env: test
spec:
  containers:
  - name: nginx
    image: nginx
    imagePullPolicy: IfNotPresent
  priorityClassName: high-priority
\end{lstlisting}

\subsection{Consequences}

\begin{itemize}
    \item (+) The system can evict lower priority services to guarantee the crucial services use their resources;
    \item (+) The resource cost is the same as before the implementation of the pattern;
    \item (+) The system can accommodate new services with high priority without impacting the resource cost;
    \item (-) The priority levels comparison depends on the externally defined configurations of the services being compared. An error in their definition can lead to possible errors in configuration;
    \item (-) If the developers define all or most of the services as a higher priority, the problem can reappear since there are no priority discrepancies to support the eviction and freedom of new resources;
    \item (-) The eviction of service can take time, leading to a time overhead in the scheduler and deployment of the crucial service; 
    \item (-) The evicted service can lose progress of its work, which can corrupt the system in case of a wrong priority labeling or lead to faulty information if not handled.
    
\end{itemize}

\subsection{Example Resolved}

\begin{figure*}[]
    \centering
    \includegraphics[width=1.1\textwidth]{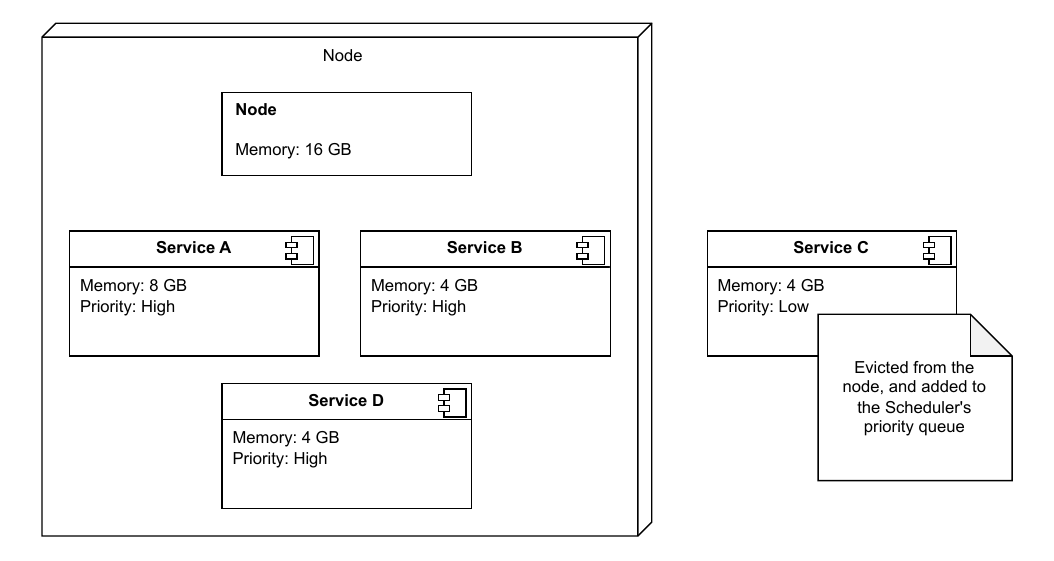}
    \caption{Preemptive Scheduling - Visual Representation of the Example Resolved}
    \label{fig:ps_res}
\end{figure*}

Considering the previous example, the 4 services would be assigned a priority level: the frontend server and API server would be assigned a priority level 1, the data processing server would be assigned a priority level 3, and the temporary service would be assigned a priority level 2. In this case, the preemptive scheduler would check the node and notice that it has all the allocated resources. Then, it would compare all the service levels and realize that the temporary service has a higher priority than the data processing server service. It would evict the data processing server, deploy the temporary server, and add the new service to the scheduler's priority queue. This can be seen in Figure \ref{fig:ps_res}.

\subsection{Known Uses}

Peloton\cite{uber_peloton}, Uber's resource management system, improves cluster utilization and reduces costs by co-locating diverse workloads, such as online services and batch jobs, on shared clusters. The system leverages preemptive scheduling to manage resources effectively.

Resource overcommitment allows Peloton to run more jobs than there are physical resources, assuming that not all jobs will require their total resource allocation simultaneously. Peloton designates lower-priority batch jobs as preemptible to prevent disruption to critical online services, which are often latency-sensitive. When the system detects that online services need more resources, it preempts (suspends or stops) these batch jobs to free up capacity for high-priority online workloads. This ensures that critical services maintain performance without interruption.

Additionally, Peloton uses disaster recovery (DR) capacity for batch jobs when the system is running normally. However, in the event of a failover, these batch jobs are preempted to ensure that the DR capacity is fully available for critical online services. Peloton can dynamically reallocate resources to essential tasks during high demand or system failover by preempting non-essential workloads. Peloton employs two types of preemption:
\begin{itemize}
    \item \textbf{Inter-resource pool preemption}: This method ensures fair distribution of resources across different resource pools by enforcing preemption policies. When necessary, it reclaims resources from specific pools according to these policies. Administrators can customize and implement different preemption strategies to suit their needs.
    \item \textbf{Intra-resource pool preemption}: This approach manages resource sharing within a single resource pool based on the priorities of the jobs. Since multiple users can use a resource pool, each running multiple jobs, conflicts can arise when one user monopolizes the pool's capacity, causing other users' jobs to be delayed and potentially miss deadlines. Additionally, when a higher-priority job arrives while lower-priority jobs are already running, the scheduler needs to free up resources for the more critical task. Intra-resource pool preemption addresses this by interrupting lower-priority jobs within the pool to make space for higher-priority ones when resources are limited.
\end{itemize}

In Linux\cite{cgroups}, preemptive scheduling allows a system to interrupt a running task (process or thread) and switch to another, typically higher-priority, task. This improves responsiveness and efficiency by ensuring critical tasks can run when needed. Linux uses preemptive multitasking, meaning it can forcibly suspend a process to run another, especially when higher-priority tasks are waiting.

Linux also supports kernel preemption, allowing system-level tasks to be interrupted for higher-priority ones. This boosts responsiveness in real-time applications. The default Linux scheduler, the Completely Fair Scheduler (CFS), balances fairness with preemption to avoid CPU monopolization.

Control groups (cgroups) let administrators allocate and limit system resources like CPU, memory, and I/O to groups of processes. This enables resource isolation—for example, prioritizing containers or user workloads. Within and across cgroups, Linux can still preempt tasks based on priority. A high-priority process in one cgroup can preempt lower-priority processes in another, depending on resource configuration. In containerized setups, this ensures critical services (e.g., a web server) get CPU time over less important jobs.

\subsection{Related Patterns}

This pattern needs a scheduler to do its job, so it would improve by applying the pattern {\sc Orchestration by Resource Offering}. An alternative to this pattern is that if the developer is not concerned with the costs and needs resources, the developer should consider scaling the system using the pattern {\sc Elasticity Manager}. The priority levels of the services can be defined using {\sc Labeling}, and the resources needed can be determined using the {\sc Resource Reserve and Limit}.

\section{Service Balancing}

\subsection{Context}

A distributed system comprises a cluster of multiple nodes. Some nodes run services, while others are empty nodes, awaiting some services to be scheduled for them. Several services are to run in nodes with various requirements and specifications. These services can already be run or be prepared on the available nodes.

Normally, during the scheduling process, a service is scheduled with a node that has the resources to run it. By placing a service in a node with more free resources, the cluster tends to balance itself. However, sometimes, problems in the system occur, such as containers going down, services requiring more resources, or nodes going down or up. These situations can lead to an imbalance in the nodes, resulting in the inefficient use of resources.

\subsection{Example}

\begin{figure}[]
    \centering
    \includegraphics[width=\textwidth]{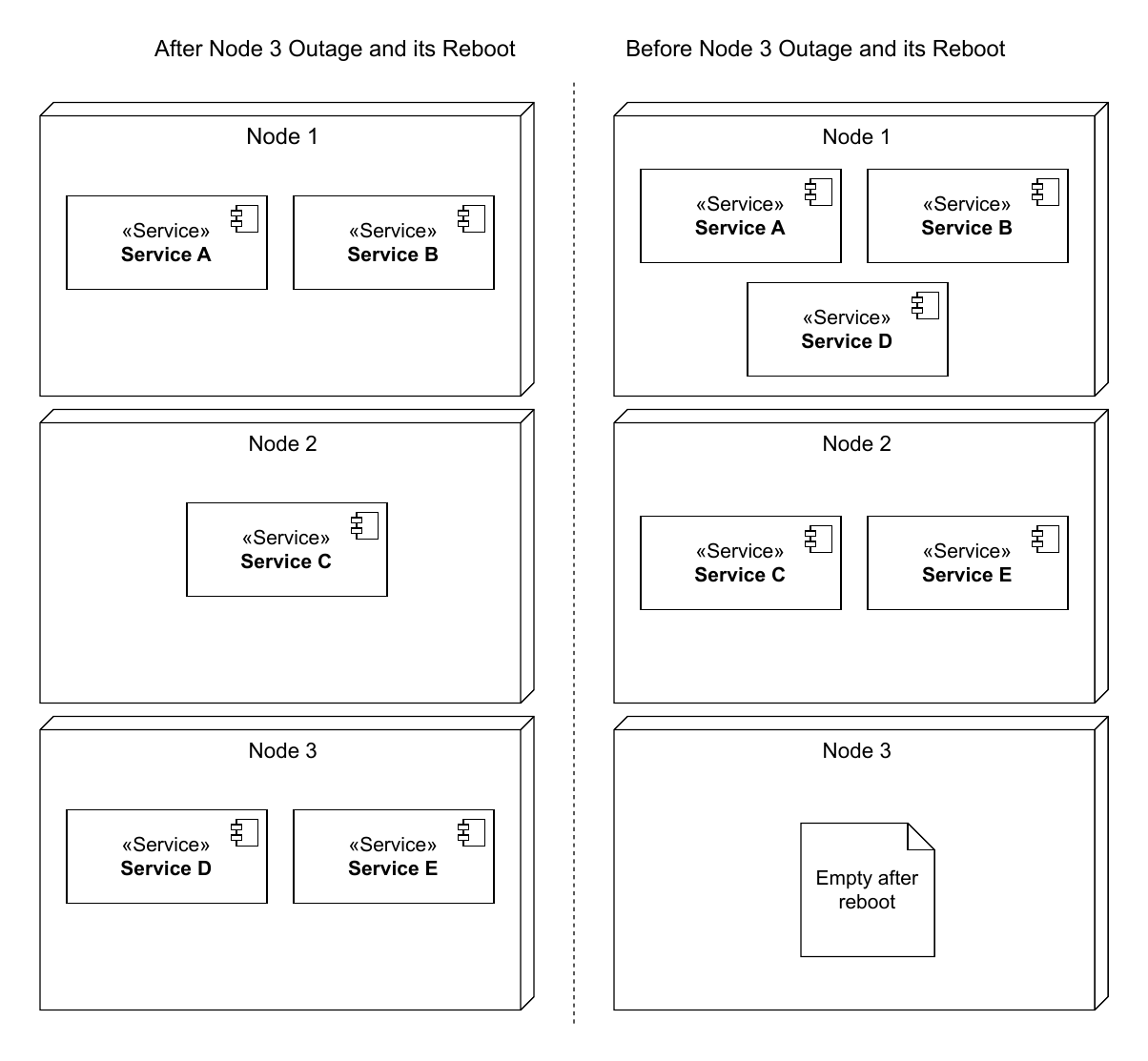}
    \caption{Service Balancing - Visual Representation of the Example}
    \label{fig:sb_1}
\end{figure}

As seen in Figure \ref{fig:sb_1}, a system built using the microservices architecture comprises 5 services: A, B, C, D, and E, and 3 nodes, each one with the same resources. Upon scheduling the services among the containers:

\begin{itemize}
    \item Node 1: service A and B
    \item Node 2: service C
    \item Node 3: service D and E
\end{itemize}

However, after an outage, node 3 is down, and in a hurry to handle fault tolerance, services D and E are scheduled into Nodes 1 and 2, leaving the system state like this:

\begin{itemize}
    \item Node 1: service A, B, and D 
    \item Node 2: service C and E
    \item Node 3: down
\end{itemize}

Nodes 1 and 2 now have almost no free resources, and after some time, node 3 is alive again and waiting for more services to be scheduled and deployed there. The system is currently imbalanced, as nodes 1 and 2 have almost fully allocated their resources, while node 3 has complete resources to be spent. 

\subsection{Problem}

How can we balance resource usage across the available nodes in a system?

\subsection{Forces}

\begin{itemize}
    \item \textbf{Resource Optimization} - An imbalanced service load leaves the system poorly using the resources, normally underutilizing them;
    \item \textbf{Stability/ Availability} - Moving around the services to obtain balance over the system can lead to uncertain consequences and unnecessary container restarts. This could also lead to the disruption of service progress if badly timed or not accounted for.
    \item \textbf{Performance of System} -  Moving around the services to obtain balance over the system can lead to a decay of the performance of the system because of the service restarts and redeployments needed;
     \item \textbf{Performance of Services} - Moving services to free nodes can lead to a performance boost for these services by having the resources they need or more; 
    \item \textbf{Fault Tolerance} - Balancing the nodes can lead to an improvement in resilience of the system when it comes to node failures;
\end{itemize}
 
\subsection{Solution}

\textit{Rebalance the services across the nodes by evicting the services considered imbalanced and scheduling them in nodes with available resources}

\vspace{0.1cm}

To effectively rebalance a cluster, three key steps must be followed: identifying the imbalance, selecting the services to rebalance, and determining the optimal time to perform the rebalancing. The primary goal of this pattern is to maximize resource optimization, even if it means temporarily sacrificing the availability of certain services.

A robust monitoring system must be in place to detect an imbalance in the cluster. Without monitoring, it's challenging to recognize when resources are unevenly distributed. This can be achieved through dashboards displaying the cluster's status or simple logging mechanisms. An imbalance typically occurs when one or more nodes are underutilized compared to others in the cluster. Once an imbalance is detected, the next step is to decide which services to reschedule. There are two main approaches:

\begin{itemize}
    \item Full Cluster Restart: Restart the entire cluster, allowing services to be rebalanced from scratch. While this can correct imbalances, it poses significant risks like service disruptions and system instability.
    \item Targeted Rescheduling: Select specific services from over-allocated nodes and restart them to redistribute resources. This approach also has risks, as restarting containers can lead to temporary instability and unpredictable outcomes.
\end{itemize}

Services can be selected for rebalancing either manually or automatically. The manual approach involves reviewing resource usage across nodes and deciding which services to relocate. Although more time-consuming, this method provides greater control and reduces the risk of performance issues. Alternatively, automation can detect significant discrepancies in resource usage between nodes and automatically select services for rebalancing. However, this approach may introduce performance and availability problems, making manual selection the safer option.

Timing is essential when rebalancing the cluster. Constant rebalancing can degrade system performance by frequently consuming resources for container evictions and rescheduling. To avoid this, rebalancing should be scheduled during maintenance windows or periods of low traffic to minimize the impact on application performance.

The rebalancing should also be done considering the loss of progression of the services that are being reallocated. This is important to reassure the resilience of the system and minimize any loss of data. With this in mind, service migrations~\footnote{\url{https://docs.oracle.com/cd/E24329\_01/web.1211/e24425/service\_migration.html\#CLUST287}} should be carefully applied. 

For large clusters, incremental rebalancing is recommended to prevent overwhelming the system. This gradual approach ensures that resources are redistributed without causing excessive strain on the cluster. Following these steps and carefully timing the rebalancing process, the cluster can be efficiently optimized while minimizing disruptions.

\subsection{Implementation}

In \textbf{Docker Swarm}~\footnote{Administer and maintain a swarm of Docker Engines: \url{https://docs.docker.com/engine/swarm/admin\_guide}}, generally, you don't need to rebalance tasks across nodes in a swarm manually. When you add a new node to the swarm, or a previously disconnected node rejoins, it won't automatically receive tasks to balance the load. This design choice avoids disrupting active services, as periodically shifting tasks to achieve balance could interfere with client applications. Instead, the swarm aims for an eventual balance with minimal user impact. New tasks or tasks from nodes that become unavailable are assigned to less busy nodes, ensuring a gradual balance over time.

If you want to force the swarm to redistribute tasks, use the \lstinline{--force} or \lstinline{-f} flag with the \lstinline{docker service update} command. This will restart service tasks and may cause temporary disruptions for client applications. If your service is set up with rolling updates, this process will help minimize interruptions.

For older versions or if you're comfortable disrupting tasks, you can manually rebalance the swarm by temporarily increasing the service scale. First, check the current scale with \lstinline{docker service inspect --pretty <servicename>}. Then, use \lstinline{docker service scale} to add more instances, which will cause tasks to be distributed to the nodes with the fewest workloads. You might need to scale up incrementally a few times to achieve a balanced load. Once the balance is satisfactory, scale the service back down to its original number of instances.

\subsection{Consequences}

\begin{itemize}
    \item (+) The node resources are being spent more optimally by occupying underutilized nodes;
    \item (+) By spreading the services across the nodes, in case of a node failure, fewer services are shut down, increasing the resistance of the system;
    \item (+) Rebalancing of the system leads to a better system performance and application availability in the long term;
    \item (-) Rebalancing of the system leads to service restart and redeployment, which worsens the distributed system performance and application availability in the short term;
    \item (-) Automatically rebalancing the services can lead to possible momentary performance and availability problems.
    \item (-) Can lead to possible loss of progress of the services migrated if not properly handled.
    
\end{itemize}
 
\subsection{Example Resolved}

\begin{figure*}[]
    \centering
    \includegraphics[width=\textwidth]{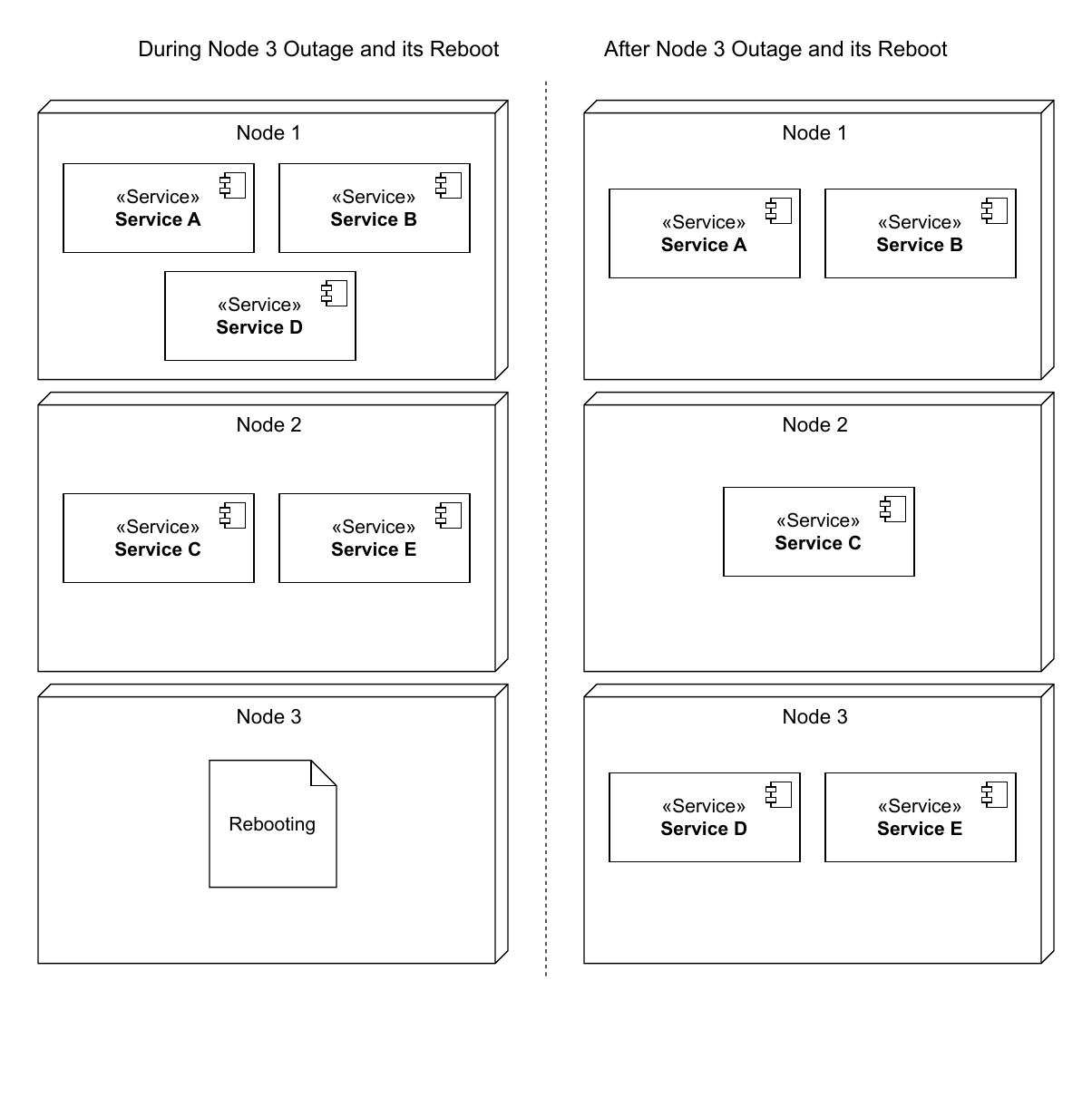}
    \caption{Preemptive Scheduling - Example Resolved}
    \label{fig:sb:res}
\end{figure*}

To address the system imbalance following the Node 3 outage, we have a clear plan of action to restore equilibrium, as we can see in Figure \ref{fig:sb:res}:

The system has become unbalanced after the Node 3 outage. With services D and E relocated to Nodes 1 and 2, these nodes are now close to full capacity, while Node 3 remains underutilized despite being fully operational. To rebalance the system, a decision must be made on which services should be moved back to Node 3. We have two approaches to consider:
\begin{itemize}
    \item Manual Selection: After reviewing the current resource usage on Nodes 1 and 2, we can manually opt to relocate services D and E back to Node 3 since they were initially stationed there. This will help in freeing up capacity on Nodes 1 and 2.
    \item Automatic Detection: An automated system can also play a role in identifying the resource usage mismatch and autonomously relocating services D and E back to Node 3, thus restoring balance. However, manual selection is favored to prevent unforeseen performance issues.
\end{itemize}

The timing of the rebalancing operation is crucial. We aim to rebalance during a maintenance window or low-traffic period to minimize disruption. This approach will reduce the impact on application performance during the restart and rescheduling of services. It is advisable to schedule the rebalancing during a quieter time and relocate services D and E back to Node 3, restoring the cluster's balance.

We can opt for a gradual rebalancing approach for larger clusters or to avoid overwhelming the system with simultaneous changes. This involves initially relocating one service, such as D, to Node 3, monitoring system stability, and later moving the second service, E.

In conclusion, the plan is to either manually or automatically relocate services D and E from Nodes 1 and 2 to Node 3, thereby effectively rebalancing the system. It is crucial to schedule this operation during a maintenance window or low-traffic period to minimize impact. By redistributing these services, Nodes 1 and 2 will free up resources, and the load will be evenly distributed across the cluster, ensuring improved system performance.

\subsection{Known Uses}

Dynamic work rebalancing in a Dataflow system from \textbf{Google Cloud}~\footnote{https://cloud.google.com/dataflow/docs/dynamic-work-rebalancing} is a method for efficiently managing tasks across different nodes or processors. Here's how it works: initially, tasks are distributed among nodes based on availability. The system constantly monitors each node's performance and workload. If some nodes become overloaded or underutilized, the system adjusts the task distribution to achieve a better balance. This involves moving tasks from busy nodes to those with more capacity.

The main benefits of dynamic work rebalancing are improved performance and resource utilization. It ensures that no single node is overwhelmed, which helps maintain smooth operation and reduces delays. Additionally, it increases resilience by adapting to changes like node failures or sudden spikes in workload. This approach also makes it easier to scale the system up or down as needed.

In parallel computing, \textbf{Load Balancing}~\footnote{\url{https://hpc-wiki.info/hpc/Load\_Balancing}} is the process of distributing computational tasks evenly across multiple processors or cores to maximize performance and efficiency. The aim is to avoid situations where some processors are overloaded while others remain idle, which leads to poor utilization and increased execution time. Load balancing strategies can be static, where tasks are assigned before execution, or dynamic, where the system reallocates tasks during runtime based on current load conditions. Efficient load balancing ensures that all processors contribute equally, resulting in faster computations and optimal use of hardware resources.

The same concept holds true in distributed systems, especially those built on microservice architectures.. Instead of balancing low-level tasks across cores, the focus is on balancing services (like containers or pods) across multiple machines or nodes. This is known as service balancing and is often managed by orchestration tools such as Kubernetes. These systems monitor resource usage (CPU, memory, etc.) and distribute or scale services to prevent any node from becoming a performance bottleneck. Much like dynamic task balancing in parallel computing, this process adapts to real-time conditions and helps maintain system health, availability, and scalability.

The core idea connecting both contexts is efficient resource distribution to avoid overload and underutilization. In parallel computing, the concern is with keeping cores busy; in distributed systems, it’s about keeping nodes balanced. Both rely on dynamic strategies that adjust to changing workloads and aim to deliver smooth, predictable performance. Ultimately, service balancing in distributed systems can be seen as a higher-level application of the same load balancing principles used in parallel computing, adapted to operate across networked environments and containerized services.

\subsection{Related Patterns}

When rebalancing the services, these services are sent to the scheduler. In this regard, we believe using the pattern {\sc Orchestration by resource offering} is an excellent option to guarantee a good scheduling process. Service balancing can use the {\sc Labelling} process to define a property not to allow a service to be scheduled.

\section{Garbage Collection}

\subsection{Context}
In container orchestration systems, applications are deployed as containers across multiple nodes in a cluster. As these systems manage the lifecycle of containers, resources such as services, images, volumes, and configurations are created and used, and eventually become obsolete. Over time, unused resources can accumulate, leading to resource wastage, performance degradation, and potential outages.

In such dynamic environments, there is a need for a mechanism that automatically cleans up resources that are no longer in use to ensure the system remains efficient and stable. This is where the garbage collection pattern in container orchestration comes into play.

\subsection{Example}

\begin{figure*}[]
    \centering
    \includegraphics[width=\textwidth]{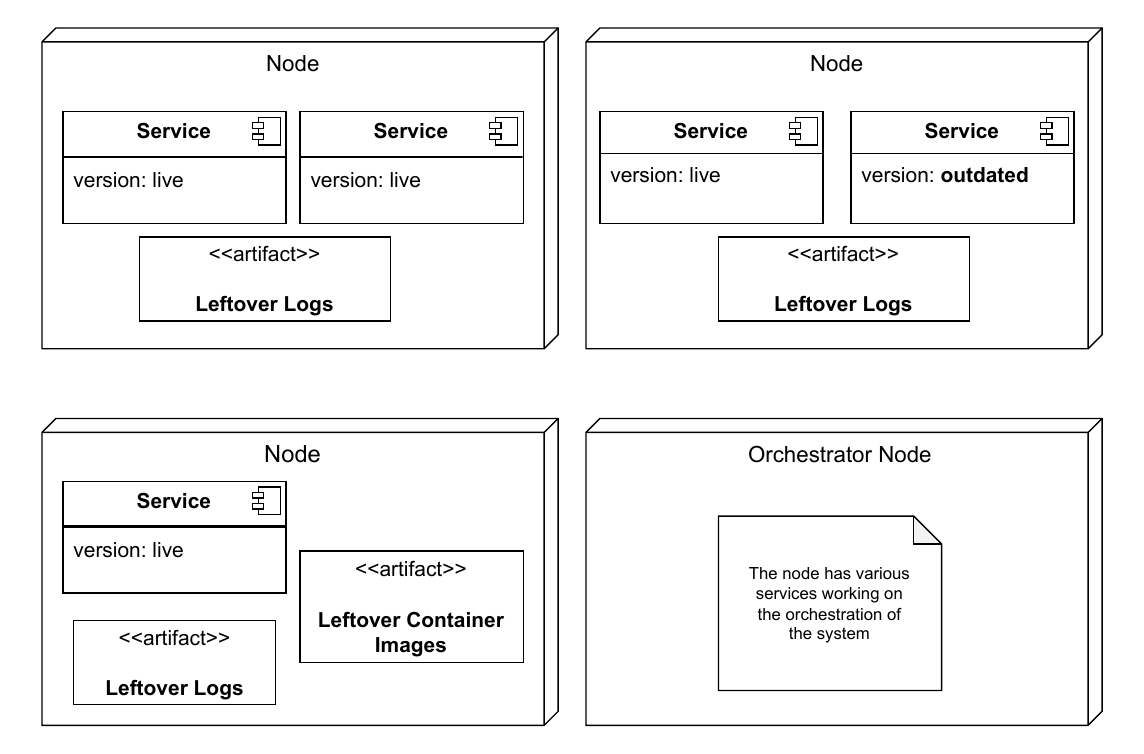}
    \caption{Garbage Collection - Visual Representation of Example}
    \label{fig:gc_ex}
\end{figure*}

A software system uses a container orchestration platform to manage its applications. The system is composed of multiple services, each responsible for different functionality. These services are deployed as sets of containers, and the system is frequently updated to add new features or address issues. The following is represented by Figure \ref{fig:gc_ex}

Over time, the accumulation of older service versions can lead to performance degradation. Residual logs and unused container images consume disk space, and the orchestration platform's API server may slow down due to the growing number of managed objects. This can result in operational challenges and a reduction in overall system performance.

\subsection{Problem}

How can a distributed system using containers prevent resource wastage, performance degradation, and stability issues caused by accumulating unused or obsolete resources?

\subsection{Forces}

\begin{itemize}
    \item \textbf{Resource Efficiency} - Ensuring that resources such as memory, CPU, and storage are efficiently utilized without being consumed by obsolete objects;
    \item \textbf{System Overhead} - Balancing the overhead of continuously monitoring and cleaning up resources with maintaining system performance;
    \item \textbf{Automation} - Automating the cleanup of unused resources to reduce operational burden
    \item \textbf{Control} - Providing sufficient control to administrators to customize or intervene in cleanup processes when necessary (e.g., setting reclaim policies, excluding certain resources from automatic deletion);
    \item \textbf{Scalability/ Complexity} - As the scale of the system grows, the cleanup mechanism must be capable of handling larger numbers of resources without becoming a bottleneck or introducing complexity that is difficult to manage;
    \item \textbf{Availability} - Ensuring that the cleanup process does not interfere with active workloads or reduce system availability, especially in environments with high uptime requirements;
    
\end{itemize}

\subsection{Solution}

\textit{Implement automated cleanup mechanisms that identify and remove unused resources while allowing customization and control to prevent resource wastage and maintain system efficiency.}

\vspace{0.1cm}

The solution is implementing a garbage collection service that identifies and removes inactive resources. This will ensure that the cluster remains clean and efficient.

For this system to work, it will need to have some components:
\begin{itemize}
    \item Lifecycle Management;
    \item Automated Deletion Mechanisms
    \item Finalizers
\end{itemize}

The Lifecycle Management system would define clear ownership and lifecycle for each object in the system, including services, containers, and volumes. Owner references would be used to track dependencies, allowing for cascading deletions, where dependent objects are automatically cleaned up when their owner is deleted. For example, a node owns the services it has. When the node is deleted, the containers are also deleted, impeding orphaned resources.

The Automated Deletion Mechanisms would clean the resources that the Lifecycle Management would detect. Some examples of functions it should have are the automatic deletion of terminated or completed services after a certain threshold to avoid accumulation, an image running system to remove unused images to free up disk space regularly, the automatic deletion of volumes that are no longer bound to any persistent claims, and the implementation of Time to Live policies for short-lived resources, such as jobs.

The Finalizers would allow custom cleanup tasks to be executed before an object is deleted, such as cascading the deletion of objects.

These are the main components a Garbage Collector should have to allow the system to clean itself and reduce its resource wastage automatically. However, garbage collection requires careful planning, like setting up lifecycle policies, monitoring resource uses, and handling edge cases that the automatic cleanup may not cover.  

\subsection{Implementation}

Kubernetes\footnote{Garbage Collection in Kubernetes: \url{https://kubernetes.io/docs/concepts/architecture/garbage-collection}} has built-in garbage collection processes for managing resources like pods, replica sets, and deployments. For instance, when a pod is terminated, Kubernetes automatically cleans up its associated resources, such as ephemeral storage and networking configurations. Additionally, Kubernetes can automatically remove old versions of images and unused volumes based on policies defined in the cluster configuration.

Kubernetes checks for and deletes objects without owner references, like the pods left behind when you delete a ReplicaSet (set of pod replicas). When you delete an object, you can control whether Kubernetes deletes the object's dependents automatically in a cascading deletion process. There are two types of cascading deletion, as follows:

\begin{itemize}
    \item \textbf{Foreground cascading deletion} - the owner object you're deleting first enters a deletion in progress state. After the owner object enters the deletion in progress state, the controller deletes the dependents. After deleting all the dependent objects, the controller deletes the owner object.
    \item \textbf{Background cascading deletion} - In background cascading deletion, the Kubernetes API server deletes the owner object immediately, and the controller cleans up the dependent objects in the background. This is Kubernetes' default garbage collection process.
\end{itemize}

You can also control how and when garbage collection deletes resources that have owner references using Kubernetes finalizers.

\subsection{Consequences}

\begin{itemize}
    \item (+) Improved resource efficiency by reducing resource consumption by automatically cleaning up unused resources and optimizing disk space by deleting older or unused objects;
    \item (+) Enhanced system performance with faster response times and reduced latency since there are fewer objects to manage;
    \item (+) Simplified maintenance since the cleanup is automatic, and this allows for easier management of resources; 
    \item (-) Risk of premature deletion if the garbage collector is not configured correctly;
    \item (-) Increase of system overhead since garbage collection mechanisms are always running;
    \item (-) Some resources may become orphaned if dependencies are not well-defined, and the garbage collector may not handle certain data types, leading to remains in the system.
\end{itemize}
 
\subsection{Example Resolved}

\begin{figure*}[]
    \centering
    \includegraphics[width=\textwidth]{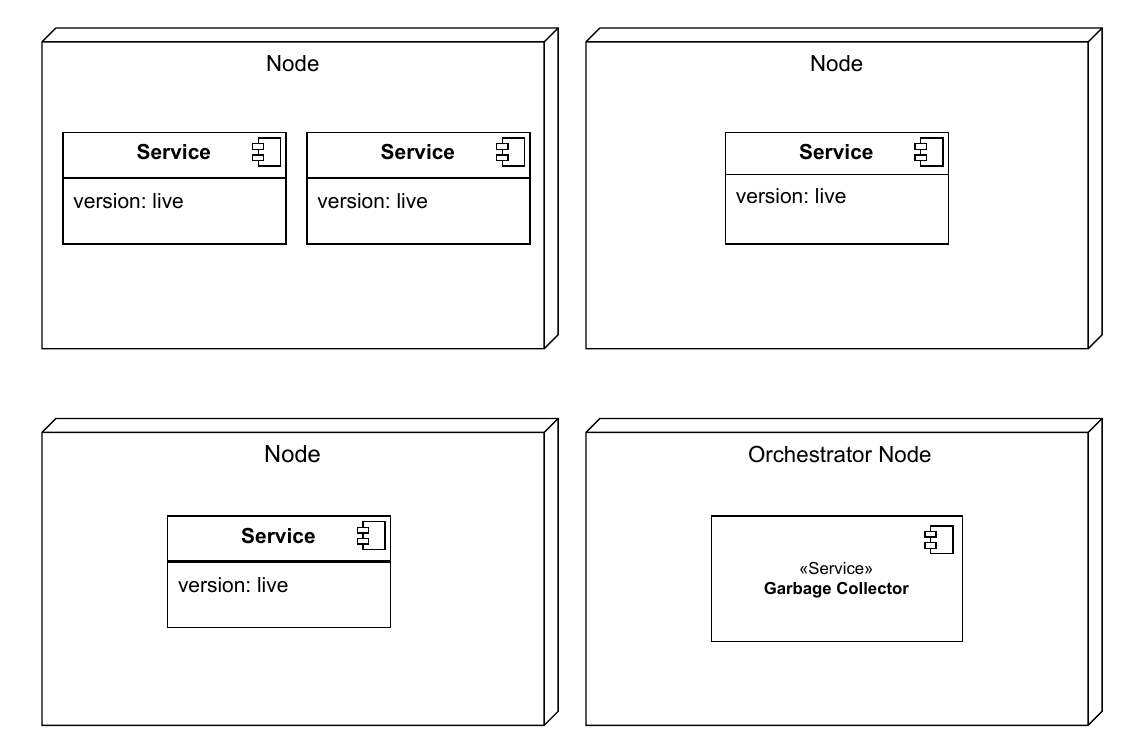}
    \caption{Garbage Collection - Visual Representation of Example Resolved}
    \label{fig:gc_res}
\end{figure*}

Returning to our previous example, we would need to implement all the functionalities described in the pattern solution. To do so, we would need to implement:

\begin{itemize}
    \item A built-in garbage collection mechanism to handle terminated services. The system would retain them for a maximum of 30 days, and after that time passes, it would delete them to free resources;
    \item To manage disk space, it would clean all images not tagged by any active deployment and not used in the last 120 days;
    \item A mechanism to delete persistent volumes associated with persistent volume claims;
    \item To control the batch jobs, it would implement a Time to Live controller. Jobs, whether completed or not, are automatically deleted after a week to ensure there is no clutter of jobs.

This would create a simple and efficient garbage collection system that manages resource wastage and prevents issues arising from accumulated objects. The resolution can be seen in Figure \ref{fig:gc_res}.

\end{itemize}

\subsection{Known Uses}

Docker Swarm\footnote{Build garbage collection: https://docs.docker.com/build/cache/garbage-collection/} uses garbage collection to clean up unused images, volumes, and networks. This helps free up disk space and ensures that the swarm operates efficiently. When containers are removed or services are updated, Docker Swarm's garbage collection mechanisms help manage and remove obsolete resources. The following are the default garbage collection policies:

\begin{lstlisting}[language=json,caption= Default garbage collection policies of Docker Swarm]
GC Policy rule#0:
        All:            false
        Filters:        type==source.local,type==exec.cachemount,type==source.git.checkout
        Keep Duration:  48h0m0s
        Keep Bytes:     512MB
GC Policy rule#1:
        All:            false
        Keep Duration:  1440h0m0s
        Keep Bytes:     26GB
GC Policy rule#2:
        All:            false
        Keep Bytes:     26GB
GC Policy rule#3:
        All:            true
        Keep Bytes:     26GB
\end{lstlisting}

\begin{itemize}
    \item \textbf{Rule 0}:  if build-cache uses more than 512MB, delete the most easily reproducible data after it has not been used for 2 days;
    \item \textbf{Rule 1}: remove any data not used for 60 days;
    \item \textbf{Rule 2}: keep the unshared build-cache under cap;
    \item \textbf{Rule 3}: if previous policies were insufficient, delete internal data to keep the build-cache under the cap.

\end{itemize}

However, developers can create rules and build custom configurations for the garbage collection process that override the default ones.

\subsection{Related Patterns}

Garbage Collection patterns are directly connected to the {\sc Containerization} pattern since containers, and their images, are part of objects that it is in charge of cleaning. Some monitoring patterns, such as {\sc Health API Checks}, are also important to this pattern since they could help the system recognize whether the object is obsolete.

\section{Conclusion}

This paper introduces three patterns related to orchestrating and optimizing the system's resources. These patterns are related to those defined in previous work and aid in solving consequence patterns in the patterns map introduced to a system. 

These methods are well-established and widely recognized as best practices within the container orchestration and Kubernetes communities. However, they have not previously been formalized as patterns. By documenting them, we aim to provide value to both application developers and those building new orchestration tools.

Additional practices will undoubtedly remain to be identified within container orchestration tools and from industry professionals. This work could also be expanded to cover other operational areas of container orchestration, such as Deployment and Scheduling, contributing to the creation of a comprehensive pattern language for container orchestration.

\section{Acknowledgments}
This work is supported by national funds, through the Operational Competitiveness and Internationalization Programme (COMPETE 2020) [Project nº 182852; Funding Reference: SIFN-01-9999-FN-182852].

We would like to thank Christian Engelmann for his valuable support during the shepherding process and for the insightful feedback he provided throughout.

We are also grateful to the authors who participated in our workshop session—Francesco Urdih, Uwe Zdun, Julia Pampus, Marcelo Nunes, Tiago Boldt Sousa, and Daniel Reis—for their constructive input, which helped refine several aspects of the work and inspired ideas that contributed to the article's development.

\bibliographystyle{splncs04}
\bibliography{references}

%\begin{credits}
%\subsubsection{\ackname} A bold run-in heading in small font size at the end of the paper is
%used for general acknowledgments, for example: This study was funded
%by X (grant number Y).

%\subsubsection{\discintname}
%It is now necessary to declare any competing interests or to specifically
%state that the authors have no competing interests. Please place the
%statement with a bold run-in heading in small font size beneath the
%(optional) acknowledgments\footnote{If EquinOCS, our proceedings submission
%system, is used, then the disclaimer can be provided directly in the system.},
%for example: The authors have no competing interests to declare that are
%relevant to the content of this article. Or: Author A has received research
%grants from Company W. Author B has received a speaker honorarium from
%Company X and owns stock in Company Y. Author C is a member of committee Z.
%\end{credits}

\end{document}